\documentclass[conference]{IEEEtran}
\newcommand{\qed}{\nobreak \ifvmode \relax \else
      \ifdim\lastskip<1.5em \hskip-\lastskip
      \hskip1.5em plus0em minus0.5em \fi \nobreak
      \vrule height0.75em width0.5em depth0.25em\fi}

\ifCLASSINFOpdf
  \usepackage[pdftex]{graphicx}
  \graphicspath{{./}{Figs/}}
  \DeclareGraphicsExtensions{.pdf,.jpeg,.png}
\else
\fi
\usepackage{amssymb}
\usepackage{amsmath}
\usepackage{mathtools}
\usepackage{bigdelim}

\newcommand{\MeijerG}[7]{G^{#1,#2}_{#3,#4} \left( \begin{smallmatrix} #5 \\ #6 \end{smallmatrix} \middle\vert #7 \right) }

\IEEEoverridecommandlockouts

\IEEEpubid{978-1-4799-3083-8/14/\$31.00 \copyright 2014 IEEE}

\begin{document}

\title{Spectrum Sensing Via Reconfigurable Antennas: Is Cooperation of Secondary Users Indispensable?}
\author{\IEEEauthorblockN{Ahmed M. Alaa, Mahmoud H. Ismail and Hazim Tawfik}
\IEEEauthorblockA{Department of Electronics and Communications Engineering,\\
Faculty of Engineering, Cairo University,
Giza 12613, Egypt.\\ Emails: \{ahmedalaa, mhismail\}@ieee.org and htawfik@eece.cu.edu.eg}}
\maketitle
\begin{abstract}
This work presents an analytical framework for characterizing the performance of cooperative and non-cooperative spectrum sensing schemes by figuring out the trade-off between the achieved diversity and coding gains in each scheme. Based on this analysis, we try to answer the fundamental question: \textit{can we dispense with SUs cooperation and still achieve an arbitrary diversity gain?} It is shown that this is indeed possible via a novel technique that can offer diversity gain for a \textit{single} SU using a \textit{single} antenna. The technique is based on the usage of a reconfigurable antenna that changes its propagation characteristics over time, thus creating an artificial temporal diversity. It is shown that the usage of reconfigurable antennas outperforms cooperative as well as non-cooperative schemes at low and high Signal-to-Noise Ratios (SNRs). Moreover, if the channel state information is available at the SU, an additional SNR gain can also be achieved.
\end{abstract}
\IEEEpeerreviewmaketitle
\section{Introduction}
Cognitive Radio (CR) is a promising technology offering enhanced spectrum efficiency via dynamic spectrum access \cite{IEEEhowto:kopka1}. In a CR network, unlicensed SUs can opportunistically occupy the unused spectrum allocated to a licensed PU. This is achieved by means of PU signal detection. Based on the sensing result, the SU is required to decide whether or not a PU exists. Energy Detection (ED) is one of the simplest spectrum sensing techniques as it can be implemented using simple hardware and does not require Channel State Information (CSI) at the SU receiver. However, the performance of ED severely degrades in fading channels \cite{IEEEhowto:kopka2}. To combat this effect, cooperative spectrum sensing schemes have been proposed to take advantage of the spatial diversity in wireless channels \cite{IEEEhowto:kopka2}, \cite{IEEEhowto:kopka3}. In cooperative spectrum sensing, hard decisions from different CR users are combined to make a global decision at a central unit known as the Fusion Center (FC).

Although cooperation among SU receivers achieves diversity gain, it encounters a significant cooperation overhead \cite{IEEEhowto:kopka4}: several decisions taken at SU terminals have to be fed back to the FC via a dedicated reporting channel; global information (including the number of cooperating SU terminals) must be provided to each SU in order to calculate the optimal detection threshold; and hard decisions taken locally at each SU cause loss of information \cite{IEEEhowto:kopka5}, which degrades the performance at low SNR. In this work, we characterize the performance of cooperative and non-cooperative spectrum sensing schemes in terms of the achieved diversity and coding gains. Based on this analysis, it is shown that cooperation is not beneficial for the whole range of SNR as the non-cooperative scheme achieves a higher coding gain. Then, we tackle the following question: \textit{can we achieve an arbitrary diversity gain without employing SU cooperation?} To answer this question, we propose a novel spectrum sensing scheme that can achieve an arbitrary diversity order for a single SU with a single antenna. The scheme is based on the usage of \textit{reconfigurable antennas}; a class of antennas capable of changing one of its characteristics (polarization, operating frequency and radiation pattern) over time \cite{IEEEhowto:kopka6, IEEEhowto:kopka7}. Hence, switching the antenna \textit{radiation state} over time will manipulate the wireless channel thus creating artificial channel fluctuations. Such fluctuations have been used in the context of interference alignment in \cite{IEEEhowto:kopka8}. Capitalizing on this property, the proposed single user spectrum sensing scheme exploits the temporal channel variations and consequently, allows dispensing with the spatial diversity achieved through cooperation. Because the proposed scheme does not involve local hard decisions, it even outperforms the cooperative scheme at low SNR. Analysis and simulations show that the proposed scheme outperforms both cooperative and non-cooperative schemes as it captures full coding and diversity gains at any SNR. Moreover, when the CSI is available at the SU receiver, an additional SNR gain, known as \textit{selection gain} will be achieved.

The rest of the paper is organized as follows: Section II presents the system model for the cooperative, non-cooperative and reconfigurable antenna spectrum sensing schemes. In Section III, we discuss the \textit{``To cooperate or not to cooperate"} tradeoff, which shows the impact of coding and diversity gains on the performance of conventional cooperative and non-cooperative schemes. Section IV presents the proposed scheme and its simulation results are depicted in Section V. Finally, we draw our conclusions in Section VI.
\IEEEpubidadjcol
\section{System Model and Notations}
We first present some notations as well as the three schemes that will be analyzed and compared in this work.
\subsection{Diversity Order and Coding Gain}
The diversity order reflects the number of independent fading channels observed in space, fequency or time. The diversity order $d$ for a performance metric $P_{*}$ with an average SNR of $\overline{\gamma}$ is defined as \cite{IEEEhowto:kopka4}
\[d = -\lim_{\overline{\gamma}\to \infty} \frac{\log P_{*}}{\log \overline{\gamma}}.\]
In our analysis, we adopt the Neyman-Pearson (NP) detection approach, thus the metric $P_{*}$ corresponds to the missed detection probability $P_{md}$. As for the coding gain, it is defined as the multiplicative factor of the average SNR in $P_{*}$ as $\overline{\gamma}$ tends to infinity. Thus, if $P_{*}$ $\asymp$ $\frac{1}{(A \overline{\gamma})^d}$ as $\overline{\gamma}$ $\mapsto$ $\infty$, the coding gain is given by $A$ and the diversity order is $d$ \cite{IEEEhowto:kopka4}, where $\asymp$ denotes asymptotic equality. Without loss of generality, we are interested in evaluating the asymptotic missed detection probability at high SNR only in order to obtain the diversity order and coding gain using the definitions above. However, both gains characterize the performance for all ranges of SNR.
\subsection{Spectrum Sensing Schemes Under Study}
\subsubsection{Non-cooperative Scheme}
A conventional non-cooperative spectrum sensing scheme involves one SU that observes $M$ samples for spectrum sensing. It is assumed that the instantaneous SNR is $\gamma$, the primary signal $i^{t h}$ sample is $S_{i}$ $\sim$ $\mathcal{CN}(0, 1)$ \cite{IEEEhowto:kopka5} and the additive white noise is $n_{i}$ $\sim$ $\mathcal{CN}(0, 1)$. Thus, the $i^{t h}$ sample received at the SU receiver is a binary hypothesis given by
\begin{equation}
\label{eqn_example}
   r_{i} = \left\{
     \begin{array}{lr}
       n_{i} \sim \mathcal{CN}(0, 1), & \ \mathcal{H}_{o} \\
       \sqrt{\gamma} S_{i} + n_{i} \sim \mathcal{CN}(0, 1 + \gamma), & \ \mathcal{H}_{1}
     \end{array},
   \right.
\end{equation}
where $\mathcal{H}_{o}$ and $\mathcal{H}_{1}$ are the null and alternative hypotheses, which denote the absence and presence of the PU, respectively. After applying such signal to an energy detector, the resulting test statistic will be $Y = \sum_{i=1}^{M} |r_{i}|^{2}$, which follows a central chi-squared distribution for both $\mathcal{H}_{o}$ and $\mathcal{H}_{1}$. The false alarm and detection probabilities are, respectively, given by \cite{IEEEhowto:kopka5}
\[P_{F} = P(Y > \lambda | \mathcal{H}_{o}) = \frac{\Gamma \left(M,\frac{\lambda}{2}\right)}{\Gamma(M)},\]
and
\begin{equation}
\label{eqn_example}
P_{D} = P(Y > \lambda | \mathcal{H}_{1}, \gamma) = \frac{\Gamma \left(M,\frac{\lambda}{2(1 + \gamma)}\right)}{\Gamma(M)},
\end{equation}
where $\lambda$ is the detection threshold, $\Gamma(\cdot,\cdot)$ is the upper incomplete gamma function and $\Gamma(\cdot)$ is the gamma function. We assume Rayleigh fading with an average SNR of $\overline{\gamma}$. The instantaneous SNR is assumed to be constant over the $M$ observed samples (slow fading) and will vary according to the exponential distribution.
\subsubsection{Cooperative Scheme}
A cooperative CR network consists of $N$ SUs, each takes a local decision on the existence of a PU and reports its decision to an FC. The FC employs an \textit{n-out-of-N} fusion rule to take a final global decision. We let $l$ be the test statistic denoting the number of votes for the existence of a PU. Hence, the conditional probability density functions (pdfs) follow a \textit{binomial distribution} \cite{IEEEhowto:kopka3}
\[P(l | \mathcal{H}_{o}) = \binom{N}{l} \hspace{1.5 mm} P_{F}^{l} \hspace{1.5 mm} (1-P_{F})^{N-l},\]
and
\begin{equation}
\label{eqn_example}
P(l | \mathcal{H}_{1}) = \binom{N}{l} \hspace{1.5 mm} \overline{P}_{D}^{l}\hspace{1.5 mm} (1-\overline{P_{D}})^{N-l},
\end{equation}
where $P_{F}$ is the local false alarm probability, and $\overline{P}_{D}$ is the local detection probability averaged over the pdf of the SNR as follows
\begin{equation}
\label{eqn_example}
\overline{P}_{D} = \int_0^{\infty} \frac{\Gamma \left(\frac{M}{2},\frac{\lambda}{2(1 + \gamma)}\right)}{\Gamma\left(\frac{M}{2}\right)} \frac{1}{\overline{\gamma}} e^{-\frac{\gamma}{\overline{\gamma}}}  \,d\gamma.
\end{equation}
Based on the fusion rule mentioned above, the global false alarm and detection probabilities $P_{F_{G}}$ and $P_{D_{G}}$ are
\[P_{F_{G}} = \sum_{l=n}^{N} \binom{N}{l} P_{F}^{l} \left(1-P_{F}\right)^{N-l},\]
and
\begin{equation}
\label{eqn_example}
P_{D_{G}} = \sum_{l=n}^{N} \binom{N}{l}  \overline{P}_{D}^{l}  \left(1-\overline{P}_{D}\right)^{N-l}.
\end{equation}

\subsubsection{Single User Spectrum Sensing Using a Reconfigurable Antenna}
In the proposed scheme, we assume a single SU that employs a reconfigurable antenna to sense the PU signal. A reconfigurable antenna can change its characteristics by dynamically changing its geometry \cite{IEEEhowto:kopka7}. Each geometrical configuration leads to a different mode of operation leading to different realizations of the perceived wireless channel. Switching between various antenna modes could be done using Microelectromechanical Switches (MEMS), Nanoelectromechanical Switches (NEMS) or solid state switches \cite{IEEEhowto:kopka8}. This would create temporal diversity for a single SU network, which can offer a gain similar to the spatial diversity gain in the cooperative scheme.

In a slow fading channel, reconfigurable antennas with $Q$ modes can offer $Q$ different channel realizations. Thus, the $i^{t h}$ sample received at the SU receiver is a binary hypothesis given by
\begin{equation}
\label{eqn_example}
   r_{i} = \left\{
     \begin{array}{lr}
       n_{i}, & \ \mathcal{H}_{o} \\
       \sqrt{\gamma_{j}} \hspace{1 mm} S_{i} + n_{i}. & \ \mathcal{H}_{1}
     \end{array}.
   \right.
\end{equation}
The conditional distributions on null and alternative hypotheses are
\begin{equation}
\label{eqn_example}
   r_{i} \sim \left\{
     \begin{array}{lr}
       \mathcal{CN}(0, 1), & \ \mathcal{H}_{o} \\
       \mathcal{CN}(0, 1 + \gamma_{j}), & \ \mathcal{H}_{1}
     \end{array},
   \right.
\end{equation}
where $\gamma_{j}$ belongs to the set of channel realizations $\{\gamma_{1}, \gamma_{2}, \cdots, \gamma_{Q}\}$. The set of $Q$ channel gains are independent identically distributed (i.i.d) Rayleigh random variables \cite{IEEEhowto:kopka7}. One can switch the antenna mode per sample as long as the switching delay is less than the sampling period.

In the next section, we introduce a new analysis for conventional cooperative and non-cooperative schemes that quantifies the missed detection probability in terms of the coding and diversity gains.

\section{To Cooperate or Not to Cooperate}
In this section, we compare the performance of both cooperative and non-cooperative schemes. Unlike the classical treatment of the spectrum sensing threshold optimization problem \cite{IEEEhowto:kopka5}, we relate the local and global thresholds ($\lambda$ and $n$) to the coding gain and diversity order. Thresholds are selected so that the diversity order is maximized with the false alarm probability being constrained to $\alpha$. This corresponds to an $\alpha$-level NP test.

\subsection{Non-cooperative Scheme Analysis}
For the non-cooperative scheme, the diversity order defined in Section II is obtained by evaluating the average detection probability defined in (4). We can rewrite the integrands in (4) in terms of the Meijer-G function $\MeijerG{m}{n}{p}{q}{a_1,\ldots,a_p}{b_1,\ldots,b_q}{z}$ [9, sec. 7.8] as $\Gamma \left(M,\frac{\lambda}{2(1+\gamma)}\right) = \MeijerG{2}{0}{1}{2}{1}{M, \hspace{0.5 mm} 0}{\frac{\lambda}{2(1+\gamma)}},$ and $e^{-\frac{\gamma}{\overline{\gamma}}} = \MeijerG{1}{0}{0}{1}{-}{0}{ \frac{\gamma}{\overline{\gamma}}}.$ With the aid of [9, Eq. 7.811.1] and after some manipulations, the integral is approximated at high SNR as
\begin{equation}
\overline{P}_{D} = \frac{2 \hspace{0.5 mm} e^{\frac{1}{\overline{\gamma}}}}{\Gamma(M)} \left( \frac{\lambda}{2 \hspace{0.5 mm} \overline{\gamma}} \right)^{\frac{M}{2}} K_{M} \left(\sqrt{\frac{2 \hspace{0.5 mm} \lambda}{\overline{\gamma}}}\right),
\end{equation}
where $K_{M}(.)$ is the $M^{th}$ order modified bessel function of the second kind. The asymptotic expansion of $K_{M}(x)$ as $x$ $\mapsto$ 0 is $K_{M}(x)$ $\asymp$ $x^{-M}\left(2^{M-1} \Gamma(M) - \frac{2^{M-3} \Gamma(M) x^{2}}{M-1} + \frac{2^{M-6} \Gamma(M) x^{4}}{(M-1)(M-2)} +\ldots\right)$ \cite{IEEEhowto:kopka9}. Note that $\sqrt{\frac{2 \lambda}{\overline{\gamma}}}$ $\mapsto$ 0 and $e^{\frac{1}{\gamma}}$ $\mapsto$ 1 as $\overline{\gamma}$ $\mapsto$ $\infty$. Hence, the asymptotic expansion of the detection probability is given by
\[\overline{P}_{D} \asymp 1 - \frac{\lambda}{2 \overline{\gamma} (M-1)} + \frac{\lambda^2}{8 \overline{\gamma}^2 (M-1) (M-2)} + \ldots.\]
Thus, at large average SNR, the first two terms dominate and $\overline{P}_{D} \approx 1 - \frac{\lambda}{2 \hspace{0.5 mm} \overline{\gamma} \hspace{0.5 mm} (M-1)}$. Also, the average missed detection probability is $\overline{P}_{md} = \frac{\lambda}{2 \hspace{0.5 mm} \overline{\gamma} \hspace{0.5 mm} (M-1)}$. As defined in Section II, the diversity order $d$ and coding gain $A$ are thus given by \[ d = -\lim_{\overline{\gamma}\to \infty} \frac{\log (\frac{\lambda}{2 \overline{\gamma} (M-1)})}{\log \overline{\gamma}} = 1,
\,\,\,\, A = \frac{M-1}{\lambda}.\]
Hence, the non-cooperative scheme achieves no diversity gain and its coding gain depends on the number of samples involved in energy detection as well as the value of the local threshold $\lambda$. For an $\alpha$-level NP test, the local threshold is decided by the value of $\alpha$ when setting $P_{F}$ = $\alpha$.

\subsection{Cooperative Scheme Analysis}
In cooperative sensing, local thresholds are set by individual SU receivers to take local hard decisions, while an integer global threshold is used by the FC to take the final decision. Based on (5), the global missed detection probability $P_{md_{G}}$ is given by
\begin{equation}
\label{eqn_example}
P_{md_{G}}= \sum_{l=0}^{n-1} \binom{N}{l} \overline{P}_{md}^{N-l}\left(1-\overline{P}_{md}\right)^{l}.
\end{equation}
It is obvious that $\overline{P}_{md}$ $\mapsto$ 0 as $\overline{\gamma}$ $\mapsto$ $\infty$. The last term in the series in (9) dominates and the asymptotic value of $P_{md_{G}}$ is
\[P_{md_{G}} \asymp \binom{N}{n-1} \hspace{1.5 mm} \bigg( \frac{\lambda}{2 \hspace{0.5 mm} \overline{\gamma} \hspace{0.5 mm} (M-1)}\bigg)^{N-n+1}. \]
Thus, the diversity order $d$ and coding gain $A$ in terms of the local and global thresholds are given by
\[d = N - n + 1, \,\,\,\, A = \binom{N}{n-1}^{\frac{1}{N-n+1}} \hspace{1.25 mm} \frac{M-1}{\lambda}.\]
It is clear that the global threshold that maximizes the diversity order is $n$ = 1, which is commonly known as the OR rule \cite{IEEEhowto:kopka3}. Hence, if only one SU votes for the presence of a primary user, the FC will adopt its decision. The local threshold $\lambda$ is chosen such that $P_{F_{G}}$ = $\alpha$. Based on the above analysis, it is shown that cooperative spectrum sensing with $N$ SU receivers can offer a diversity order of $N$. The larger $N$ is, the higher the diversity order is, but the more information is lost due to local hard decisions at each SU. In the low SNR range, information loss due to the poor coding gain is more critical and we do not benefit from multiuser diversity. Thus, for a fixed total energy constraint, it is better not to cooperate when the SNR is low as assigning the total energy to a single SU leads to a better detection performance.

To demonstrate the tradeoff between coding and diversity gains, we compare a cooperative network with $N$ SU terminals and $M$ samples per terminal with a non-cooperative network with a single SU and $NM$ samples. The total sensed energy is kept constant in both cases to ensure a fair comparison. Let the local thresholds for multiple and single users be $\lambda_{MU}$ and $\lambda_{SU}$, respectively. The coding gain would be $\frac{M-1}{\lambda_{MU}}$ in the cooperative scheme and $\frac{NM-1}{\lambda_{SU}}$ in the non-cooperative scheme. Thus, the coding gain of the non-cooperative scheme is boosted by a factor of $N$. This factor is slightly reduced as $\lambda_{MU}$ and $\lambda_{SU}$ are not generally equal.
\begin{figure}[t]
\centering
\includegraphics[width=3.25 in]{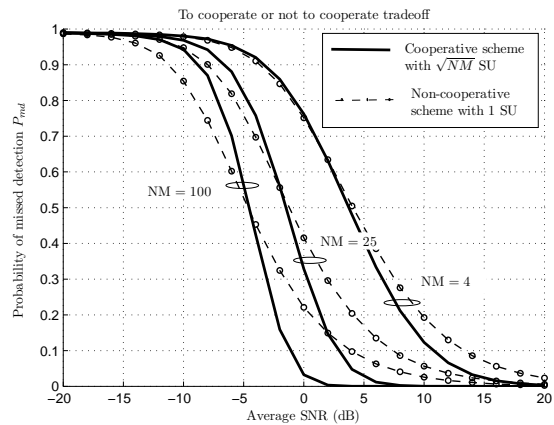}
\caption{To cooperate or not to cooperate tradeoff.}
\label{fig_sim}
\end{figure}
Figure 1 depicts the tradeoff under study. Simulations are carried out for cooperative and non-cooperative schemes and the missed detection probability is plotted versus the average SNR. The $NM$ product is fixed for both schemes and is set to 4, 25 and 100. For each value, the cooperative scheme employs $\sqrt{NM}$ SU terminals and $\sqrt{NM}$ samples per terminal. On the other hand, the non-cooperative scheme employs 1 SU and $NM$ samples. By applying an NP test and setting $\alpha$ = 0.01, it is found that at $NM$ = 100, the non-cooperative scheme outperforms the cooperative scheme by 3 dB at low SNR. Thus, it is better not to cooperate if the operating SNR is less than $-5$ dB. This SNR gain is reduced in the $NM$ = 25 scenario and nearly vanishes when $NM$ = 4. On the other hand, the cooperative scheme offers large gains in the high SNR range. For instance, at $P_{md}$ = 0.03 and $NM$ = 100, cooperation offer a gain of about 7 dB due to the multiuser diversity. The larger $N$ is, the more gain we get at high SNR. It is worth mentioning that the relatively long local sensing period in the non-cooperative scheme is compensated by the time spent by SUs to report their local decisions to the FC in the cooperative scheme. In the next section, we present an analysis for the proposed spectrum sensing scheme and quantify the achieved diversity order when using a reconfigurable antenna at the SU receiver.

\section{Spectrum Sensing via Reconfigurable Antennas}
In this section, we present a spectrum sensing scheme that uses a reconfigurable antenna with the following advantages: first of all, the proposed scheme captures the full coding and diversity gains; it behaves like the non-cooperative scheme at low SNR and matches the cooperative scheme at high SNR. Moreover, it overcomes the space limitation problem that inhibits the usage of multiple antennas. In addition to that, unlike multiple antenna systems, only one RF chain is needed. Also, as mentioned earlier, the availability of the CSI does not benefit the conventional non-cooperative schemes. Contrarily, the proposed scheme can benefit from the CSI to boost the achieved coding gain. Finally, diversity is achieved with no cooperation overhead, which involves setting up a dedicated reporting channel, feeding back information from the FC to SU terminals and maintaining synchronization between SU devices.

We investigate two basic schemes for spectrum sensing using a reconfigurable antenna: a \textit{state switching} scheme (when the CSI is unknown) and a \textit{state selection} scheme (when the CSI is available). The pattern of antenna modes selected by the SU over time is referred to as the \textit{antenna mode signature}.

\subsection{Spectrum Sensing via State Switching}
In this scenario, the SU has a set of $Q$ random channel realizations $\{\gamma_{1}, \gamma_{2}, \ldots, \gamma_{Q}\}$. For sensing over $M$ samples, the SU is able to switch to a new channel realization every specific number of samples. In this case, the de facto channel is a fast fading channel with a coherence time that depends on the rate of antenna state switching. This fast fading channel is artificially created by varying the antenna radiation pattern or polarization over time, and does not involve the motion of primary or secondary users. In order to allow for all of the $Q$ channel realizations to be employed within the sensing period, assume that $M$ is greater than $Q$, and $l_{j}$ is the number of samples assigned to one of the possible $Q$ channel realizations. Based on the signal model presented in Section II, the test statistic resulting at the output of the energy detector can be written as
\[Y = \sum_{j=1}^{Q} (1+\gamma_{j}) \hspace{0.5 mm} x_{j},\]
where $\gamma_{j}$ is one of $Q$ independent channel realizations and $x_{j}$ is a chi-square distributed random variable with $2 l_{j}$ degrees of freedom. Thus, the probability of missed detection will simply be the cumulative density function (CDF) of the linear combination of chi-square random variables. An extremely accurate approximation for the CDF of the sum of weighted chi-square random variables was proposed in \cite{IEEEhowto:kopka10}. Based on Eqs. (20-23) in \cite{IEEEhowto:kopka10}, the probability of missed detection will be given by
\[P_{md} = \min\{H(w), G(w)\}, \,\,\, w = \frac{\lambda}{M + \sum_{j=1}^{Q} l_{j} \gamma_{j}},\]
\[G(w) = \sum_{j=1}^{2M} w \hspace{0.5 mm} \frac{1+\gamma_{j}}{\lambda} \times \frac{\Upsilon\left(\frac{\lambda}{2w(1+\gamma_{j})}, \frac{\lambda}{1+\gamma_{j}}\right)}{\Gamma\left(\frac{\lambda}{2w(1+\gamma_{j})}\right)},\]
\[H(w) = \frac{\Upsilon \left(M, \frac{\lambda}{\sqrt[M]{\prod_{j=1}^{Q}(1+\gamma_{j})^{l_{j}}}}\right)}{\Gamma(M)}\]
where $\Upsilon(\cdot, \cdot)$ is the lower incomplete gamma function. Given that $\Upsilon(M, x)$ $\asymp$ $\frac{x^M}{M}$ as $\overline{\gamma}$ $\to$ $\infty$ \cite{IEEEhowto:kopka9}, the asymptotic values of $H(w)$ and $G(w)$ are $\frac{\lambda^M}{\Gamma(M+1)\prod_{j=1}^{Q} (1+\gamma_{j})^{l_{j}}}$ and $\sum_{j=1}^{2M} w \hspace{0.5 mm} \frac{1+\gamma_{j}}{\lambda}$, respectively, which means that $\min\{G(w), H(w)\} = H(w)$ at large SNR. Thus, we can calculate the diversity order based on $P_{md}$ = $H(w)$. The asymptotic missed detection probability will then be given by
\begin{equation}
P_{md}(\gamma_{1},\ldots,\gamma_{Q}) \asymp \frac{\lambda^M}{\Gamma(M+1)\prod_{j=1}^{Q} (1+\gamma_{j})^{l_{j}}}.
\end{equation}
By averaging the missed detection probability in (10) over the pdf of $Q$ independent Rayleigh channel realizations we get
\[\overline{P}_{md} = \frac{\lambda^M}{\Gamma(M+1)} \int_{\gamma_{1}=0}^{\infty} \int_{\gamma_{2}=0}^{\infty} \ldots \int_{\gamma_{Q}=0}^{\infty} \frac{1}{\prod_{j=1}^{Q} (1+\gamma_{j})^{l_{j}}} \times \]
\[\frac{1}{\overline{\gamma}^Q} e^{\frac{-\sum_{j=1}^{Q} \gamma_{j}}{\overline{\gamma}}} d \gamma_{1} d \gamma_{2} \ldots d \gamma_{Q},\]
which can be reduced to
\begin{equation}
\overline{P}_{md} = \frac{\lambda^M}{\Gamma(M+1)} \prod_{j=1}^{Q} \int_{\gamma_{j}=0}^{\infty} \frac{1}{ (1+\gamma_{j})^{l_{j}}} \frac{1}{\overline{\gamma}} e^{\frac{- \gamma_{j}}{\overline{\gamma}}} d \gamma_{j}.
\end{equation}
It can be easily shown that the integral in (11) is given by
\[\overline{P}_{md} = \frac{\lambda^M}{\Gamma(M+1)} \prod_{j=1}^{Q} \overline{\gamma}^{-l_{j}} e^{\frac{1}{\overline{\gamma}}} \Gamma \Big(1-l_{j}, \frac{1}{\overline{\gamma}}\Big).\]
At large SNR, $e^{\frac{1}{\overline{\gamma}}}$ $\to$ 1 and $\Gamma(1-l_{j}, \frac{1}{\overline{\gamma}})$ $\asymp$ $\frac{\overline{\gamma}^{l_{j}-1}}{l_{j}-1}$ yielding
\begin{equation}
\overline{P}_{md} \asymp \frac{\lambda^M}{\Gamma(M+1)} \times \frac{1}{\prod_{j=1}^{Q} (l_{j}-1) \hspace{1 mm} \overline{\gamma}^{Q}}.
\end{equation}
Designing the optimum antenna modes signature is related to the choice of the number of samples $l_{j}$ associated to an antenna realization $\gamma_{j}$. It is obvious from (12) that minimizing the missed detection probability is achieved by maximizing the quantity $\prod_{j=1}^{Q} (l_{j}-1)$. We can obtain the optimum values of the $l_{j}$'s via a simple \textit{Lagrange optimization problem} as
\[l_{1} = l_{2} = \ldots = l_{Q} = \lfloor \frac{M}{Q} \rfloor,\]
where $\lfloor.\rfloor$ is the flooring operator. Thus, the optimum antenna switching pattern is achieved by changing the antenna radiation mode every $\lfloor \frac{M}{Q} \rfloor$ samples. Note that this result is intuitive as all channel realizations are independent and identically distributed, which means that the optimal antenna mode switching pattern is reached when every mode is employed for an equal time interval during the sensing period. From (12), the achieved diversity order is $d = Q$. It is obvious that if the number of samples is less than the number of antenna states, only $M$ channel realizations can be employed during the sensing period. Thus, the diversity order will be generally given by $d = \min\{M, Q\}$. Due to the approximations involved in the proposed analysis, the asymptotic expression in (12) can not reflect an accurate coding gain. However, it is obvious that the amount of energy sensed by this scheme is the same as that sensed by the non-cooperative scheme\footnote{As shown in Section III, the coding gain depends only on the detection threshold and number of samples. Both quantities are equal for the non-cooperative and sensing via reconfigurable antenna schemes.}.

\subsection{Spectrum Sensing via State Selection}
In the non-cooperative scheme, knowledge of the CSI at the SU can not add any coding or diversity gains to the detection performance. In the proposed scheme, however, the CSI will be utilized to \textit{select} the ``best" antenna mode (the mode with largest channel gain) rather than \textit{switch} the antenna modes over time. This resembles \textit{selection combining} in multiple antenna systems. Thus, an SNR gain is obtained that is termed as a \textit{selection gain}. The pdf of the maximum gain of $Q$ Rayleigh distributed channels is given by
\[
f_{\gamma_{max}}(\gamma_{max}) = \frac{Q}{\overline{\gamma}} e^{\frac{-\gamma_{max}}{\overline{\gamma}}}(1-e^{\frac{-\gamma_{max}}{\overline{\gamma}}})^{Q-1}.
\]
In order to simplify the analysis, we focus on the dominant fading density at large $\overline{\gamma}$, which can be written as \cite{IEEEhowto:kopka11}
\[ f_{\gamma_{max}}(\gamma_{max}) \approx \frac{Q}{\overline{\gamma}^{Q}} e^{\frac{-\gamma_{max}}{\overline{\gamma}}} \gamma_{max}^{Q-1},\]
and the probability of missed detection as a function of the instantaneous channel gain is thus\cite{IEEEhowto:kopka5}
\[P_{md} = \frac{\Upsilon \big(M, \frac{\lambda}{2(1+\gamma_{max})}\big)}{\Gamma(M)}.\]
Consequently, the average missed detection probability is given by
\begin{equation}
\overline{P}_{md} = \frac{Q}{\overline{\gamma}^{Q}} \int_{0}^{\infty} \frac{\Upsilon \left(M, \frac{\lambda}{2(1+\gamma_{max})}\right)}{\Gamma(M)}  e^{\frac{-\gamma_{max}}{\overline{\gamma}}} \gamma_{max}^{Q-1} d\gamma_{max}.
\end{equation}
For simplicity, assume that $1+\gamma_{max}$ $\approx$ $\gamma_{max}$ for large SNR ranges at which the diversity gain is calculated. The integrands in (13) can be represented in terms of the Meijer-G function as
\[\overline{P}_{md} = \frac{Q}{\Gamma(M) \overline{\gamma}^{Q}} \int_{0}^{\infty}\gamma_{max}^{Q-1} e^{-\frac{\gamma_{max}}{\overline{\gamma}}} \MeijerG{1}{1}{1}{2}{1}{M,  0}{\frac{\lambda}{2  \gamma_{max}}} d \gamma_{max}.\]
Using the property $\MeijerG{m}{n}{p}{q}{a_1,\ldots,a_p}{b_1,\ldots,b_q}{z}$ = $\MeijerG{n}{m}{q}{p}{1-b_1,\ldots,1-b_q}{1-a_1,\ldots,1-a_p}{z^{-1}}$ followed by
[9, Eq. 7.813], the average missed detection probability can be represented by
\[\overline{P}_{md} = \frac{\mathcal{K}_{1}}{\overline{\gamma}^Q} \hspace{1 mm} {}_{1}F_{2}(Q; Q+1, -M+Q+1; \frac{\lambda}{2 \hspace{0.5 mm} \overline{\gamma}})\]
\begin{equation}
+ \frac{\mathcal{K}_{2}}{\overline{\gamma}^M} \hspace{1 mm} {}_{1}F_{2}(M; M+1, -M+Q+1; \frac{\lambda}{2 \hspace{0.5 mm} \overline{\gamma}}),
\end{equation}
where $\mathcal{K}_{1}$ and $\mathcal{K}_{2}$ are constants, ${}_{p}F_{q}(a_{1},...,a_{p}; b_{1},...,b_{q}; z)$ is the generalized hypergeometric function, which tends to 1 as $z$ $\to$ 0. Thus, it can be easily concluded that the diversity order of the state selection scheme is given by $d = \min\{M, Q\}$. Selecting the best channel state every sensing period offers a coding gain that we refer to as the \textit{selection gain}. This gain could be quantified as the ratio between the average SNR in the state selection scheme relative to the state switching scheme 
\begin{equation}
\mbox{Selection gain} = \frac{E\{\gamma_{max}\}}{E\{\gamma\}} = H_{Q},
\end{equation}
where $E\{.\}$ is the expectation operator and $H_{Q}$ is the $Q^{th}$ harmonic number defined as $H_{Q} = 1 + \frac{1}{2}+\frac{1}{3}+...+\frac{1}{Q}$. For large number of antenna states, the selection gain tends to
\[\mbox{Selection gain} \approx \log(Q)-\psi(1),\]
where $-\psi(1)\approx 0.577$ is the $\textit{Euler-Mascheroni}$ constant. Thus, the coding gain obtained from state selection grows logarithmically with the number of antenna states.

\section{Simulation results}
We are interested in investigating the impact of using spectrum sensing via a reconfigurable antenna on the trade-off between the diversity and coding gains. Based on an $\alpha$-level NP test with $\alpha$ = 0.05 \footnote{Any choice for the value of $\alpha$ will not affect our conclusions. In an NP test, there is a one-to-one mapping between $\alpha$ and $\lambda$. Thus, the value of $\alpha$ affects the coding gain of $P_{md}$ in a similar manner for all schemes.}, simulations for non-cooperative, cooperative and reconfigurable antenna schemes are shown in Fig. 2. In all schemes, the total energy constraint (or equivalently, the total number of sensed samples) is 100. For the cooperative scheme, the number of SUs is $N$ = 10 and the number of samples per SU is $M$ = 10, while for the non-cooperative scheme, the number of samples is 100. For the reconfigurable antenna scheme, the number of antenna states $Q$ = 10 and $M$ = 100. It can be observed that at SNRs less than $-6$ dB, the non-cooperative scheme has an SNR gain of about 2.5 dB relative to the cooperative scheme. At high SNR, the cooperative scheme benefits from the diversity gain and clearly outperforms the single user case. For the reconfigurable antenna scheme with no CSI, the induced temporal fluctuations in the wireless channel creates a diversity order of $Q$ = 10. Besides, at low SNR, the state switching scheme performance nearly coincides with that of the non-cooperative one. If the CSI is available at the SU, state selection can be applied resulting in an additional coding gain of $10 \log_{10}(H_{10})$ = 4.7 dB. Thus, the state selection scheme offers a significant performance boost compared to all other schemes.

The coding gain obtained by state selection can be utilized in a different manner. The spectrum sensing period in any cognitive radio scheme is a crucial parameter that decides the achievable SU throughput. One can convert the achieved coding gain into a throughput gain by reducing the number of sensed samples while keeping the performance of the state selection the same as the state switching scheme. This is achieved if the number of sensing samples is reduced to $M'$ = $\max\{\lceil\frac{M}{H_{Q}}\rceil, Q\}$. Fig. 3 depicts the performance of state selection scheme with $M'$ = $\lceil\frac{100}{H_{10}}\rceil$ = 33. The $P_{md}$ curve for the state selection scheme shows that it is possible for the proposed technique to outperform the non-cooperative and cooperative schemes while saving more than 66\% of the sensing period, which can boost the achievable SU throughput.
\begin{figure}[!t]
\centering
\includegraphics[width=3.25 in]{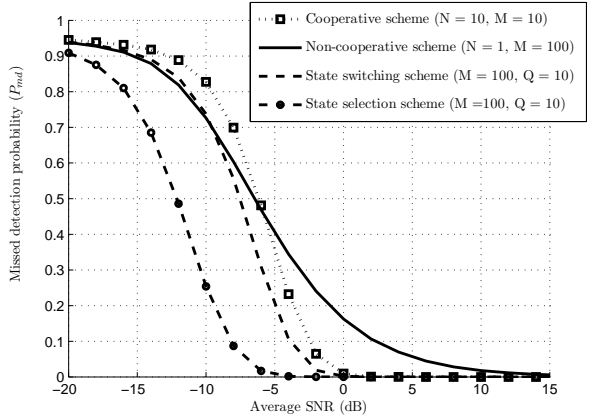}
\caption{Probability of missed detection versus the average SNR for various schemes.}
\label{fig_sim}
\end{figure}
\begin{figure}[!t]
\centering
\includegraphics[width=3.25 in]{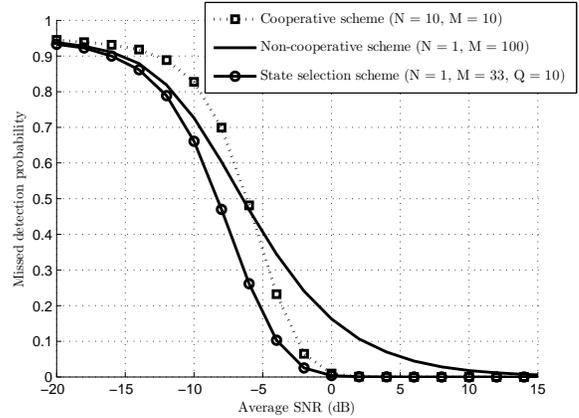}
\caption{Performance of state selection scheme with a reduced sensing period.}
\label{fig_sim}
\end{figure}
\section{Conclusions}
In this work, we presented an analysis of cooperative and non-cooperative spectrum sensing schemes identifying the achieved diversity order and coding gain in each. In light of this analysis, we presented a novel spectrum sensing scheme that uses a reconfigurable antenna to create an artificial temporal diversity by varying the operating antenna mode over time. The proposed scheme was able to dispense with SU cooperation that offers arbitrary spatial diversity but demands a significant overhead. Based on the availability of the CSI at the SU, two approaches were presented as alternatives to cooperative spectrum sensing: state switching and state selection. While the state switching scheme outperforms both non-cooperative scheme and cooperative schemes for any SNR, the state selection scheme offers an additional selection gain that can be transformed into either a better SNR or a higher throughput gain.


\begin{thebibliography}{1}
\bibitem{IEEEhowto:kopka1}
S.~Haykin, ``Cognitive radio: brain-empowered wireless communications," \emph{IEEE Journal on Selected Areas in Communications}, vol. 23, pp. 201-220, Feb. 2005.
\bibitem{IEEEhowto:kopka2}
A.~Ghasemi and E.~Sousa, ``Collaborative spectrum sensing for opportunistic access in fading environments," \emph{First IEEE International Symposium on New Frontiers in Dynamic Spectrum Access Networks}, pp. 131 – 136, Nov. 2005.
\bibitem{IEEEhowto:kopka3}
Wei Zhang, R.~K.~Mallik and K.~Ben Letaief, ``Cooperative Spectrum Sensing Optimization in Cognitive Radio Networks," \emph{Proceedings of IEEE International Conference on Communications (ICC'08)}, pp. 411 – 3415, May 2008.
\bibitem{IEEEhowto:kopka4}
D. Duan, L. Yang, and J. C. Principe,``Cooperative Diversity of Spectrum Sensing for Cognitive Radio Systems," \emph{IEEE Transactions on Signal Processing}, vol. 58, pp. 3218-3227, Jun. 2010.
\bibitem{IEEEhowto:kopka5}
Jun Ma, Guodong Zhao  and  Ye Li, ``Soft Combination and Detection for Cooperative Spectrum Sensing in Cognitive Radio Networks," \emph{IEEE Transactions on Wireless Communications}, vol. 7, pp. 4502-4507, Nov. 2008.
\bibitem{IEEEhowto:kopka6}
Yaxing Cai and Zhengwei Du,``A Novel Pattern Reconfigurable Antenna Array for Diversity Systems," \emph{IEEE Antennas and Wireless Propagation Letters,}, vol. 8, pp. 1227-1230, 2009.
\bibitem{IEEEhowto:kopka7}
P. A. Martin, P. J. Smith, and R. Murch, ``Improving Space-Time Code Performance in Slow Fading Channels using Reconfigurable Antennas," \emph{IEEE Communications Letters}, vol. 16, pp. 494-497, Apr. 2012.
\bibitem{IEEEhowto:kopka8}
T. Gou, C. Wang, and S. A. Jafar, ``Aiming Perfectly in the Dark-Blind Interference Alignment Through Staggered Antenna Switching," \emph{IEEE Transactions on Signal Processing}, vol. 59, pp. 2734-2744, Jun. 2011.
\bibitem{IEEEhowto:kopka9}
Alan Jeffrey and Daniel Zwillinger, ``Table of Integrals, Series, and Products" \emph{Academic Press}, 2000.
\bibitem{IEEEhowto:kopka10}
J. Hillenbrand, T. A. Weiss, and F. K. Jondral,``Calculation of Detection and False Alarm Probabilities in Spectrum Pooling Systems," \emph{IEEE Communications Letters}, vol. 9, pp. 349-351, Apr. 2005.
\bibitem{IEEEhowto:kopka11}
Ning Kong, ``Performance Comparison among Conventional Selection Combining, Optimum Selection Combining and Maximal Ratio Combining," \emph{Proceedings of IEEE International Conference on Communications (ICC'09), Dresden, Germany}, pp. 1-6, June 2009.

\end{thebibliography}
\end{document}